\documentclass[10pt]{article}

\usepackage[utf8x]{inputenc}
\usepackage{graphicx} 
\usepackage{epsf}
\usepackage{a4}
\usepackage{booktabs}
\usepackage{caption}
\usepackage{textgreek}
\usepackage{tabularx}
\usepackage{amsmath}
\usepackage{amssymb}
\usepackage{authblk}
\usepackage{xcolor}
\usepackage{tikz}
\usepackage{color}
\usepackage{cite}
\usepackage{multirow}

\usetikzlibrary{shapes,arrows,snakes}

\usepackage[font=footnotesize,labelfont=bf]{caption}

\usepackage{geometry} 
\usepackage{graphicx} 

\usepackage{booktabs} 
\usepackage{array} 
\usepackage{paralist} 
\usepackage{verbatim} 
\usepackage{subfig} 

\usepackage{fancyhdr} 
\usepackage{sectsty}
\allsectionsfont{\sffamily\mdseries\upshape} 

\usepackage[nottoc,notlof,notlot]{tocbibind}
\usepackage[titles,subfigure]{tocloft} 
\usepackage{amsmath}
\usepackage{relsize}

\def\beq{\begin{equation}}
\def\eeq{\end{equation}}
\def\bea{\begin{eqnarray}}
\def\eea{\end{eqnarray}}
\def\EQ{\begin{equation}}
\def\EN{\end{equation}}

\def\be{\begin{equation}}
\def\ee{\end{equation}}

\definecolor{mygreen}{rgb}{.0,.5,.1}

\begin{document}

\title{How many phases nucleate in the \\
 bidimensional Potts model?}
\author{Federico Corberi$^1$,  
Leticia F. Cugliandolo$^{2,3}$, Marco Esposito$^4$, \\
Onofrio Mazzarisi$^{1,5}$ and Marco Picco$^2$}
\affil{\small $^1$\textit{Dipartimento di Fisica E. R. Caianiello and INFN, gruppo collegato di Salerno, Universit\`a di Salerno, via Giovanni Paolo II 132, 84084 Fisciano (SA), Italy}\\
$^2$\textit{Sorbonne Universit\'e, CNRS UMR 7589, Laboratoire de Physique Th\'eorique et Hautes Energies,\\
4 Place Jussieu, 75252 Paris Cedex 05, France}\\
$^3$\textit{Institut Universitaire de France, 1, rue Descartes, 75231 Paris Cedex 05, France}\\
$^4$\textit{Imperial College London, Chain Building, 7 Imperial College Rd, Kensington, London, SW7 2AZ, UK}\\
$^5$\textit{Max Planck Institute for Mathematics in the Sciences, Inselstra$\beta$e 22, 04103 Leipzig Germany}
}

\maketitle
\thispagestyle{empty} 

\abstract{We study the kinetics of the two-dimensional $q>4$-state Potts model after a shallow quench
slightly below the critical temperature and above the pseudo spinodal. We use numerical methods and we focus on intermediate 
values of $q$, $4<q\leq 100$. We show that, initially, the system evolves as if it were
quenched to the critical temperature. The further decay from the metastable state occurs by nucleation 
of $k$ out of the $q$ possible phases. For a given quench temperature, $k$ is a logarithmically
increasing function of the system size. This unusual finite size dependence is a consequence of 
a scaling symmetry underlying the nucleation phenomenon for these parameters.}

\newpage


\section{Introduction}

When a control parameter is changed in such a way that a thermodynamical system 
undergoes a first-order phase transition one observes hysteresis and metastability.
These phenomena are widespread in nature~\cite{Debenedetti1996} and, particularly, 
in many areas of physics~\cite{Rikvold1995,Gunter1993,Gunter1994,Debenedetti1996,Debenedetti2001}. 
Besides, metastability plays a prominent role also in biological systems such as,
for instance, proteins and nucleic acids~\cite{Thirumalai2011,Volker2008}. 

The simplest example of metastability is, perhaps, a uniaxial ferromagnet in an external field, whose
paradigmatic modelisation is the Ising model.
Upon  field reversal, provided that its magnitude is sufficiently small and the temperature is 
subcritical, the magnetisation remains 
at the pre-reversal value for a certain time, the lifetime of the metastable state, before
flipping to the new equilibrium value.
The phenomenon can be ascribed to the competition between surface tension and bulk
energy and is accounted for, at least at a simple semi-quantitative level, by   
classical nucleation theory~\cite{Debenedetti1996,Frenkel1946,Laaksonen1995,Chaikin1995}: a {\it nucleus} of the equilibrium phase
in a metastable sea is unstable unless its size exceeds a critical value. If it is not the case
it disappears. Then, in this picture, the lifetime of the metastable state is the time needed
to nucleate, by thermal fluctuations, a critical nucleus (notice that  in this context {\it critical}
does not refer to the criticality of the state at $T_c$). Elaborating on these ideas, 
more refined
theories of nucleation~\cite{Becker1935,Burton1977,Penrose1979,Capoccia1974,Penrose1971,Stillinger1995,Corti1995,Langer1967,Langer1968,Langer1980,Gunther1980,Schmitz2013} 
have been developed, and they describe the phenomenon with good accuracy. 

In the previous example, the metastable and equilibrium states are associated to
the two ergodic components which characterise the system al low temperatures. 
However, the situation is not as well  understood in systems with more ergodic components,
such as those described at the simplest level by the ferromagnetic Potts model~\cite{Potts52}. This model is defined in terms
of lattice discrete variables taking $q\ge 2$ possible values, sometimes called colours, and the interaction  
favours equal colour on nearest neighbour sites.  For $q=2$,  the Ising model is recovered. 

In the absence of external fields, the Potts model undergoes a phase transition between a
high temperature disordered phase and a low temperature, symmetry broken, ordered one
at a $q$-dependent critical temperature $T_c(q)$. In two dimensions, the character of the transition changes crossing
$q=4$~\cite{Wu82,Baxter82}: for $q\le 4$ it is continuous (second order), while for $q>4$ it is discontinuous (first 
order)  and metastability is found. Specifically, upon quenching from 
$T>T_c$ to $T\lesssim T_c$ a finite-size system remains for some time in the disordered state
before starting the evolution towards the final equilibrium state. The usual interpretation 
is that the metastable state is 
the (single) ergodic component at $T>T_c$ while the target equilibrium state is one of the 
$q$ symmetry-related ergodic components at $T<T_c$. Hence, at variance with the 
field-driven transition in the Ising model, nucleation occurs towards a $q$-degenerate
final state, and one speaks of {\it multi-nucleation}. Another important difference is that
the transition is temperature driven.

To determine the lifetime of the metastable state, its dependence on the system size, the dynamic
escape from it, and the nature of the nucleation process is a difficult and much debated 
issue~\cite{Gunton83,Binder87,Oxtoby92,Kelton10}. In particular, 
it was argued~\cite{Meunier00,Ferrero09,Ferrero11,Berganza2014} that both the lifetime of the metastable state and 
temperature range below $T_c$ where it takes place shrink when the system size increases. 
This intriguing feature, which has no analog in the discontinuous field-driven transition of the Ising model, 
poses yet unresolved questions on the nature of the metastable state in the thermodynamic
limit.  In a previous paper~\cite{usprevious}, some of us 
addressed this problem analytically with a perturbation scheme valid for large $q$.
It turns out that, for $q\to \infty$, a long lived metastable state exists for quenches to final temperatures 
$T>T_\ell$, where $T_\ell \to T_c/2$. In~\cite{usprevious} a rather precise description
of the metastable state was given, both from the microscopic and thermodynamic points of view.
However, a similar characterisation of the escape kinetics is still missing for finite $q$. 
Specifically, while the nature of the metastable state can be well described in such an
analytical framework, its lifetime (for finite $q$) has not  been determined  yet
and its very existence in the thermodynamic limit cannot be proved. 
Related to this issue,  the order in which the two limits of large $q$ and
large system size should be taken remains to be clarified. 

In this paper we study the process whereby the disordered metastable state is run away and how the 
target equilibrium state is approached in the square lattice 
Potts model with $q> 4$, after shallow quenches from high to low temperatures
close to $T_c$. 
We use Monte Carlo numerical simulations with the Metropolis rule 
for different values of $q$ and linear system sizes~$L$. We choose $q$ values which 
are larger than $4$, $q=9, 24, 100$, but not too large so that we can observe nucleation in a 
relatively wide temperature interval below $T_c$, see~\cite{usprevious}.
In a companion paper by some of us~\cite{ChCuPi21} the related issue of deep quenches below the pseudo-spinodal 
of the same model for much larger $q$ values, closer to the ideal $q\to\infty$ limit, 
is tackled and the full parameter dependence of the curvature-driven growing length in the coarsening regime
is determined.

Our main result is the characterisation of the multi-nucleation process.
We show that, soon after the quench, the system evolves as if it were at the critical temperature.
Later, multi-nucleation takes over. We prove that for a given system size, the number $k$ of phases that nucleate
is a logarithmically increasing function of $L$, see Eq.~(\ref{num_phases}). 
In addition, we also study the dependence of $k$ on $q$ and $T$.
Phases which do not nucleate disappear from the system immediately after the
metastable state is escaped. Those which nucleate start competing, resulting in 
a coarsening process~\cite{Lifshitz62,Ferrero07,Olejarz13,Denholm19,Denholm20,ChCuPi21,Glazier90,Sanders2007,Ibanez07b,Petri08,Loureiro10,Loureiro12} 
which leads, at some point, to the progressive elimination
of the less represented colours until symmetry is definitively broken, one single colour
survives, and equilibrium is attained (at not too low temperatures; otherwise, blocked states of the kind 
discussed in~\cite{Lifshitz62,Ferrero07,Olejarz13,Denholm19,Denholm20,ChCuPi21} can stop the evolution). 

The paper is organised as follows. In Sec.~\ref{sec:model} we introduce the Potts model and 
its dynamics. We also define the relevant observables that will be further considered.
In Sec.~\ref{theprocess} we provide a general description of the whole dynamical process,
from the instant of the quench, to the escape from metastability and further on.
Section~\ref{secmulti} is the core of the article and contains a thorough discussion of the multi-nucleation process.
Finally, Sec.~\ref{sec:discussion} concludes the paper. 

\section{Model and observables}
\label{sec:model}

The Hamiltonian of the Potts model~\cite{Potts52} reads
\begin{equation}
H[\{s_i\}] = - J \sum_{\langle ij \rangle} \delta_{s_is_j}
\; ,
\end{equation}
where $J>0$ is a coupling constant and $\langle ij\rangle$ are nearest-neighbour couples on a two-dimensional
lattice of linear dimension $L$ which we take to be a square one. 
$\delta_{ab}$ is the Kronecker delta and $s_i$ is an integer variable ranging from 1 to $q \ge 2$. 
We consider periodic boundary conditions.
In the sum we count each bond once, and for this geometry  the energy is 
bounded between $-2JL^2$ and $0$.
The critical temperature is exactly known in $d=2$~\cite{Potts52} 
\begin{equation}
k_BT_c (q) = \frac{J}{\ln\left(1+\sqrt{q} \right)}
\label{eq:Tc}
\end{equation} 
and the phase transition is second (first) order for $q\le 4$ ($q>4$)~\cite{Baxter73}.
Notice that for finite systems of size $L$ there are corrections to the value of $T_c(q)$ given for infinite 
size in Eq.~(\ref{eq:Tc}) according to  $(T_c(L,q)-T_c(q))/T_c(q) \simeq L^{-d}$~\cite{Challa86}. 
Henceforth we will set $k_B=J=1$.  

We consider the dynamics where single spins are updated according to a random sequence
with a probability $w(s,s')$ to change from an initial value $s$ to a final one $s'$. We will use the Metropolis
form $w(s,s')=\min [1,\exp(-\Delta E/T)]$, where $\Delta E$ is the energy variation associated
to the flip. A unit time, or Montecarlo step, is elapsed after $L^2$ attempted moves.

In the following we will consider a protocol in which the system, initially prepared in an infinite
temperature equilibrium state, is suddenly quenched at time $t=0$ to a low temperature 
$T <T_c$. The dynamical evolution has been studied in previous 
works~\cite{ChCuPi21,Lifshitz62,Ferrero07,Ibanez07b,Glazier90,Sanders2007,Petri08,Loureiro10,Loureiro12,Olejarz13,Denholm19,Denholm20}
mainly focusing on the coarsening regime. 
Here we are more interested in metastability and multi-nucleation, which clearly arise 
at $T\lesssim T_c$. 

After the quench the system relaxes to a low energy state and the energy density 
$e (t)=\langle H[\{s_i\}](t) \rangle/L^2$ approaches the final equilibrium value $e(\infty)$.
Here and in the following the non equilibrium average $\langle \dots \rangle$ is 
taken over thermal histories and initial conditions. In order to quantify the relaxation,
therefore, we will consider the energy density excess 
\be
\Delta e(t)=e(t)-e(\infty).
\ee 

In a state with well formed domains, connected regions with the same spin state,  the excess energy is stored on domain walls, and
$\Delta e(t)$ is proportional to the density of interfacial spins. Since for non fractal aggregates
this quantity is, in turn, inversely proportional to the typical size of the domains, from
$\Delta e(t)$ one can infer such characteristic length as
\be
R(t)=\Delta e(t)^{-1}.
\label{rvse}
\ee
We will discuss below another determination of a typical length scale, that should behave in a
similar way.

A fundamental quantity, usually considered in coarsening processes, is the equal time spin-spin
correlation function. Specifically, we define the quantity
\begin{equation}
C_{ij}(r,t) = 
\left. 
\dfrac{\sum\limits_{n=1}^q [ \langle \delta_{s_i(t), n} \delta_{s_j(t), n} \rangle - 
\langle \delta_{s_i(t), n} \rangle \langle \delta_{s_j(t), n} \rangle ]
}
{
\left(\sum\limits_{n=1}^q   [ \langle \delta^2_{s_i(t), n} \rangle - 
\langle \delta_{s_i(t), n} \rangle^2 ]\right)^{1/2}
\left(\sum\limits_{n=1}^q [ \langle \delta^2_{s_j(t), n} \rangle - 
\langle \delta_{s_j(t), n} \rangle^2 ]\right)^{1/2}
}
\right|_{|\vec r_i - \vec r_j| = r}
\!\!\!\!\!\!\! .
\label{correl}
\end{equation}
Notice that, due to isotropy and homogeneity, $C_{ij}(r,t)$ only depends
on the distance $r=\vert \vec r \vert$, where $\vec r$ is the vector joining $i$ and $j$,
and should be independent of $i$ and $j$. Hence, further on we will denote it
$C(r,t)$ and, enforcing this symmetry, we 
will rather compute the spatial average of the quantity in Eq.~(\ref{correl}), namely 
$L^{-2}\sum _{ij} C_{ij}(r,t)$. In this way,  we will improve the statistics.  
It is clear that, on the one hand, at $r=0$ (namely $i=j$) $C(r=0,t)=1$ and, on the other hand, at very large distance,
$C(r\to\infty,t)=0$ because of the expected factorisation of the 
first term in the numerator. 
Defining 
\begin{equation}
\sigma ^2_n(t) =\langle \delta^2_{s_i(t), n} \rangle - 
\langle \delta_{s_i(t), n} \rangle^2 ,
\label{var_boolean}
\end{equation}
Eq.~(\ref{correl}) can be re-written as 
\be
C(r,t)=\frac{\sum\limits_{n=1}^q \sigma^2_n(t)
C_n(r,t)}
{\sum\limits_{n=1}^q \sigma^2_n(t)},
\label{cvscrest}
\ee
where 
\be
C_n(r,t)=\dfrac{  \langle \delta _{s_i(t),n}\delta _{s_{j}(t),n}\rangle -\left 
\langle \delta _{s_i(t),n}\right \rangle \langle \delta _{s_{j}(t),n}\rangle }
{\sigma^2_n(t)}
\label{correlp}
\ee
is the correlation function restricted to the colour $n$.
Equation~(\ref{cvscrest}) transparently expresses the fact that the unrestricted correlation is the 
weighted average of the restricted ones, the weights being the variances of the
Boolean variables associated to $s_i$, see Eq.~(\ref{var_boolean}).  

When dynamical scaling holds, $C(r,t)$ takes the form
\be
C(r,t)=g\left [\frac{r}{{\cal R}(t)}\right ] \; ,
\label{cscal}
\ee
where $g$ is a scaling function and ${\cal R}(t)$ a characteristic size with the meaning of
the typical domain's  linear dimension. 
 From the correlation function one can extract such a size in different ways.
 For instance, one
 can use the momenta as
 \be
 {\cal R}(t)=\left [\frac{\sum _r r^m\, C(r,t)}{\sum _r C(r,t)}\right ]^{\frac{1}{m}}
 \; .
 \ee
In case of dynamical scaling, determinations with different values of $m$ provide
proportional results (however large $m$ values are numerically problematic since
the noisy large-$r$ tails of $C(r,t)$ are heavily weighted).
Similarly, one can use the half-height width to define ${\cal R}(t)$ as
\be
C({\cal R}(t),t)=\frac{1}{2}
\; .
\label{half_height}
\ee
In the following we will use this determination which is easier. 
Analogously, the typical size of domains of a specific colour ${\cal R}_n(t)$ can be defined
replacing $C$ with $C_n$ in Eq.~(\ref{half_height}):
\be
C_n({\cal R}_n(t),t)=\frac{1}{2}
\; .
\label{half_height_p}
\ee

Let us mention that  Eq.~(\ref{half_height_p}) cannot be used to determine ${\cal R}_n$ in two situations: 
i) The corresponding colour is not present in the system. In this case $C_n\equiv 0$,  Eq.~(\ref{half_height_p}) 
loses its meaning, and  ${\cal R}_n$ could be taken to be identically zero.
ii) The colour has invaded the whole system. Also in this case one has $C_n\equiv 0$, because of the 
subtraction term in Eq.~(\ref{correlp}), and one can assume that
 the size of the domain of colour $n$ is the system size $L$. 
In general, immediately before one of the two situations i), ii) occur, one finds a very small value of ${\cal R}_n$. 
Hence it must be kept in mind that finding a small value of 
${\cal R}_n$ does not necessarily mean that the corresponding colour is present only in small domains, but it could 
as well be that it has flooded the system. Which case applies must be ascertained differently, 
for instance, by visual inspection of the configurations.
This will be important further on. 
 
Labelling of the $q$ possible values of the spins will be done in the following in a dynamical way: 
at any time we order the colours according to their abundances, from the most 
to the least represented, and label them with $n$ increasing from $1$ to $q$. This method is useful,
particularly in the late stage of the dynamics  where, as we will discuss, a clear hierarchy of abundances sets in.
We have checked that some exchanges in the ordering occur during evolution
for the $q$ values we use here but these are rare. Roughly speaking, a hierarchy of colours is installed when the 
phases nucleate.

\section{The dynamical process} 
\label{theprocess}

\begin{figure}[h!]
$\;$ \hspace{1.5cm} (a) \hspace{6cm} (b) \hspace{2cm}
   \begin{center}
    \includegraphics[width=0.44\textwidth]{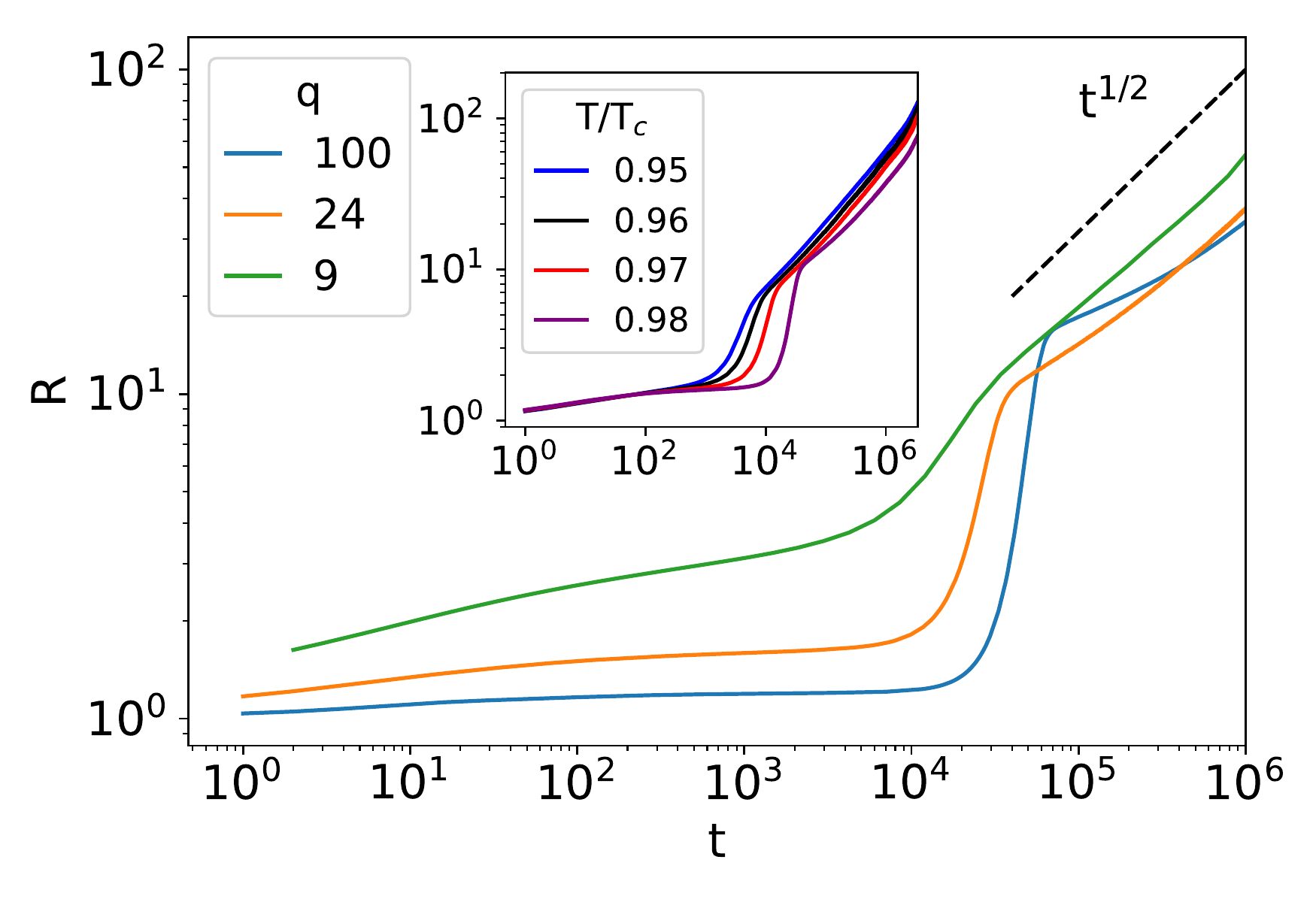}
    \includegraphics[width=0.44\textwidth]{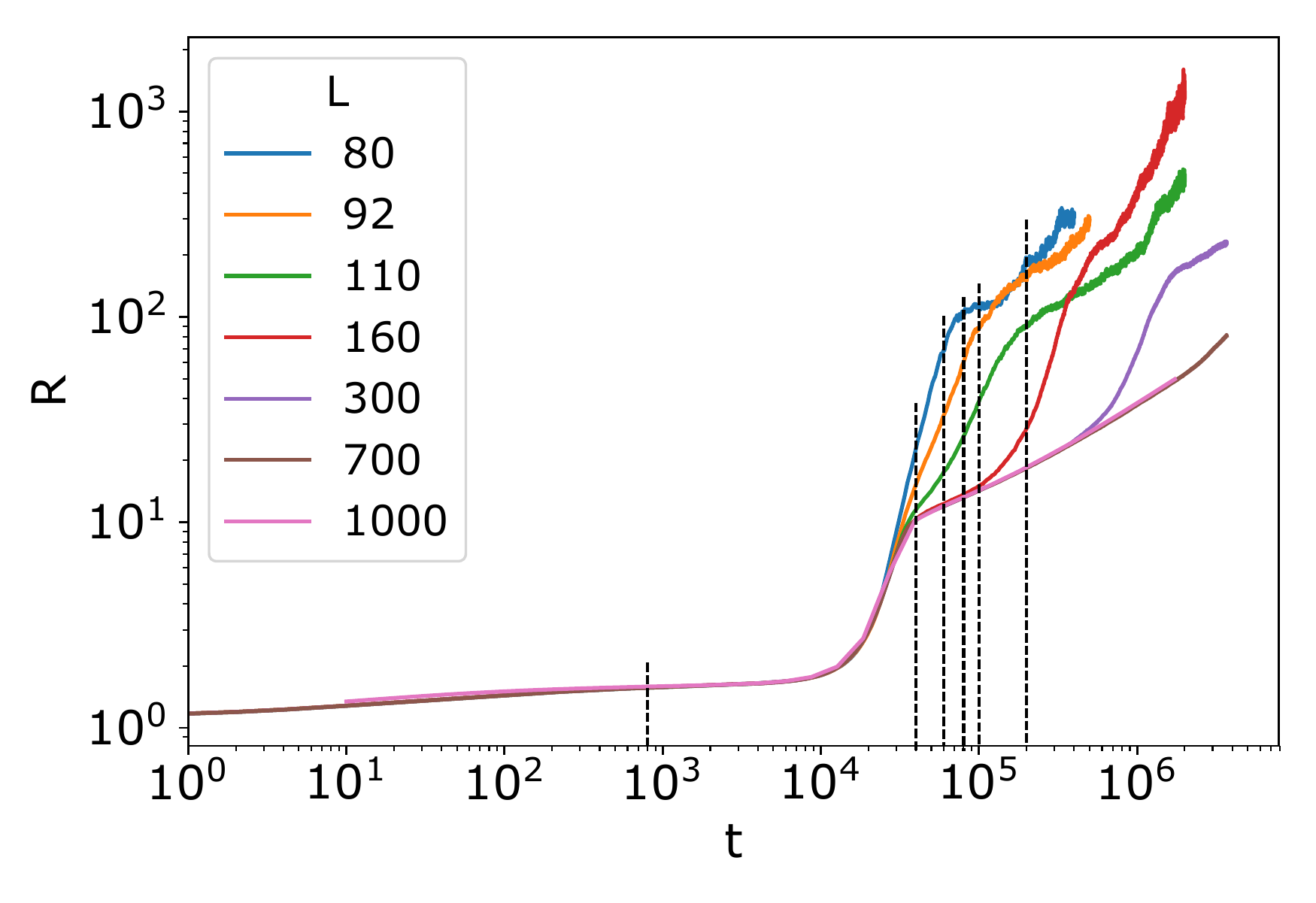}
    \end{center}   
    \caption{(a) The time evolution of growing length $R$ for $q=9,24,100$ (see the key). The linear system size is $L=700$. 
    The dashed segment shows the asymptotic curvature driven growth $t^{1/2}$. In the inset the $T/T_c$ dependence of $R$ in the case $q=24$ and 
    $L=700$ is shown. (b) $R(t)$ for $q=24$ and different linear system sizes 
    $L=80, 92, 110, 160, 300, 700, 1000$ (the last two curves superimpose).
    The dashed vertical segments indicate the times at which the snapshots of Fig.~\ref{fig:q24Snapshot} were taken.} 
\label{fig:e}
\end{figure}

\begin{figure}[h!]
   \begin{center}
   \vspace{0.25cm}
    \includegraphics[width=.9\textwidth]{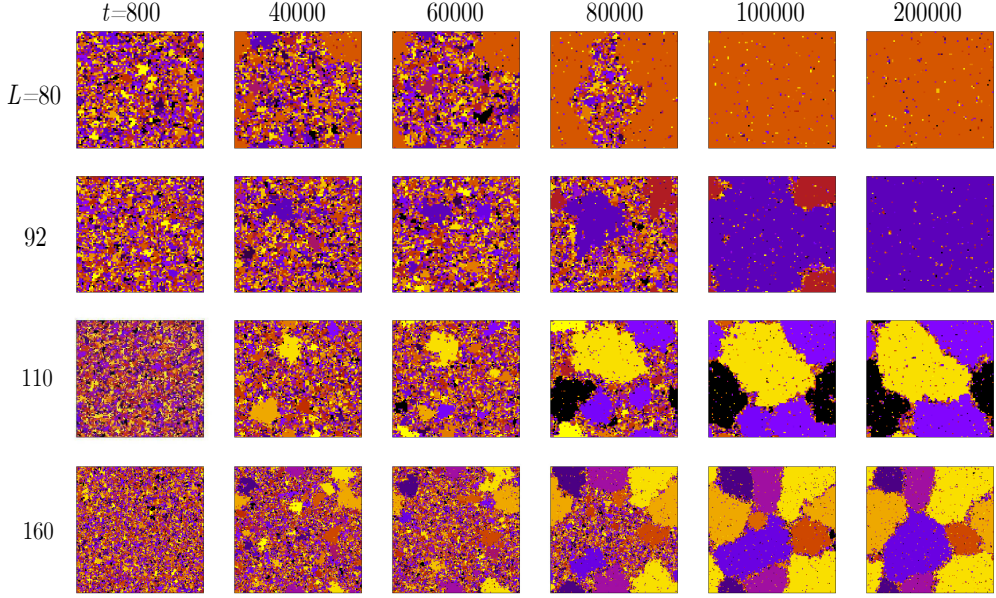}
    \end{center}
    \caption{Snapshots of a typical evolution of the $q = 24$ Potts model quenched to 
    $T=0.98 \cdot T_c$ for different lattice sizes. It is possible to appreciate that the number of phases
    that nucleate increases with the linear size of the system: $k=1$ for $L=80$, $k=2$ for $L=90$,  
    $k=3$ for $L=110$ and $k=7$ for $L= 160$, all at $t=10^5$.}
    \label{fig:q24Snapshot}   
\end{figure}

In the following we will discuss the results of the numerical simulations of the Potts model introduced above. 
We will mainly consider three $q$ values,
$q=9$, 24, 100, as paradigms of small, intermediate and large $q$ behaviors. Other $q$ values have also been 
studied and we sometimes include them in the presentation.
These systems are quenched to final temperatures close to the critical one, $T_c(q)$, 
in order to appreciate the metastable state and its lifetime. In most of our simulations we set
$T=0.9912\cdot T_c(q)$ for $q=9$, $T=0.98\cdot T_c(q)$ for $q=24$, and $T=0.95\cdot T_c(q)$ for $q=100$. 
These choices are motivated by the requirement to have a comparable nucleation time $\tau (q,T)$ (see discussion
below, and Fig.~\ref{fig:e}) for the three reference values of $q$. Notice that fixing $\tau(q,T)$ does not correspond to fixing 
$(T_c(q)-T)/T_c(q)$. Let us also remind that, as discussed below Eq.~(\ref{eq:Tc}), $T_c(q)$ is the 
critical temperature of the infinite system. In order to investigate the finite-size effects, a central issue of this paper, we
will used systems in the range of sizes $L\in [200, 1000]$ (only for qualitative illustrations, such as the snapshots in 
Fig.~\ref{fig:q24Snapshot}, we consider systems as small as $L=80$). Comparing the value of the equilibrium coherence
length $\xi$ at $T_c(q)$ given in~\cite{Buddenoir93} with the system sizes we use, one concludes that  
$L/\xi$ is at least 13 (in the worst case $L=200$ with $q=9$). Therefore our studies are expected to be 
basically free from the usual equilibrium finite size effects occurring when $L$ is smaller or comparable to $\xi$.
This, however, does not exclude other finite size effect of non-equilibrium nature, which are indeed a central
point of this article, as we will explain below.

\vspace{0.25cm}

A first qualitative description of the kinetic process after a quench stems from the behavior of $R(t)$,
the typical length scale extracted from the analysis of the excess energy as in Eq.~(\ref{rvse}),
which is shown in Fig.~\ref{fig:e}. 
With some quantitative differences, the same
kind of pattern is observed for any value of $q$ (see also~\cite{Ferrero07,Petri08} and \cite{Corberi19} where this quantity, for more $T$ and $q$ values, respectively, 
was shown). 
After a short transient, one can clearly identify 
three extended regimes. The first one 
is associated to $R$ growing slowly or staying approximately constant,
$R(t)\simeq R_m$. 
This plateau, which is flatter and longer as $T\to T_c^-$ at fixed $q$ or as $q\to \infty$ at fixed $(T-T_c)/T_c$, 
is a clear manifestation of metastability.  
In this time lag the kinetics is useless because the system is confined in a local free energy minimum
and the trapping barrier is not yet jumped over. In order to do it, critical nuclei, namely ordered domains 
of a sufficiently large size, must develop. However, this is an activated process which requires a certain time
$\tau (q,T)$, the nucleation time. At $t\ll \tau (q,T)$ the system is still in a rather disordered
state, visually not too different from the initial one, see the snapshots 
taken at $t=800$ in Fig.~\ref{fig:q24Snapshot}.
Some small ordered domains can be spotted but they are too tiny to start nucleation.

\begin{figure}[h!]
$\;$ \hspace{0.85cm} (a) \hspace{4.15cm} (b) \hspace{4.15cm} (c)
\vspace{-0.25cm}
   \begin{center}
   \vspace{0.25cm}
\includegraphics[width=1.\textwidth]{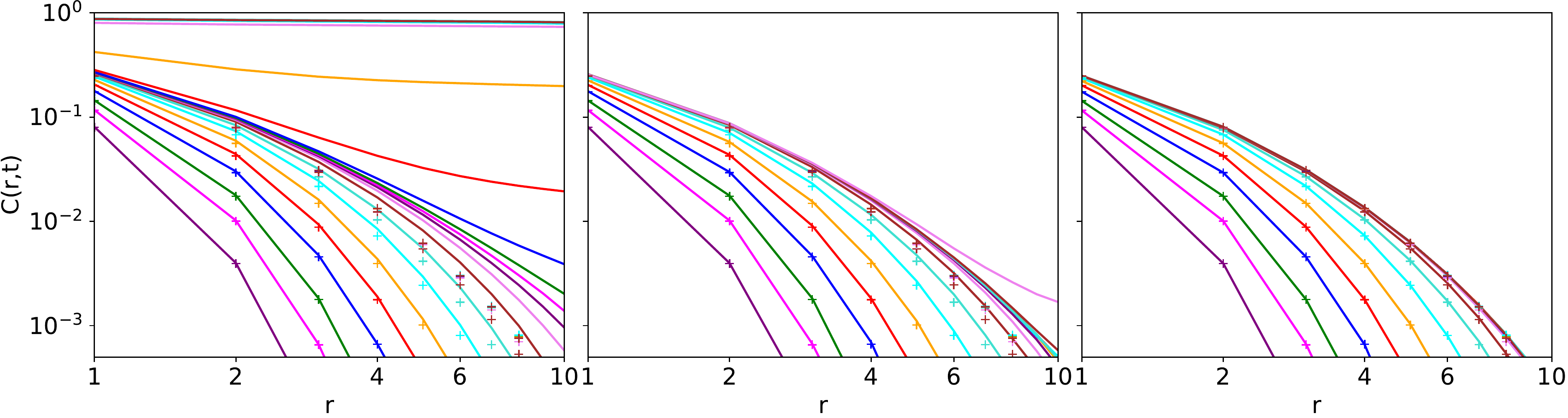}
    \end{center}
    \caption{
    The space-time correlation function $C(r,t)$ is plotted with continuous lines against 
    $r$ for a quench of the model with $q=24$, $L=700$, from
    infinite temperature to $T=0.99 \cdot T_c$ (a), $T=0.995 \cdot T_c$ (b),
    $T=0.999  \cdot T_c$ (c). Different curves correspond to different times 
    exponentially spaced (from bottom to
    top $t= 2,\ 4, \ 7, \ 15, \ 31,\ 64,\ 137,\ 291,\ 618,\ 1316,\ 2802,\ 5968,\ 12709,\ 27066,\ 57642,\ 122762,\ 261451,\ 556822,\ 1185886$).
    The crosses refer to a quench to $T=T_c$.
    See the text for a discussion.}
    \label{fig:corr}   
\end{figure}

Actually, this metastable state is 
{\it similar} to the equilibrium one at $T_c$, in a sense that we are going to specify below. We illustrate this point
by means of the correlation function defined in Eq.~(\ref{correl}). In Fig.~\ref{fig:corr} this quantity is plotted 
against $r$ during the evolution after a quench. Let us focus, to start with, on panel (a), where curves for 
a quench to $T=0.99 \cdot T_c$ (continuous lines) are compared to those, taken at the same times, for a quench to $T=T_c$ (dashed ones). 
In the latter case one sees that, as time goes on, the correlation extends to larger values of $r$ until 
at $t\simeq t_{eq}(T_c)\simeq 10^3$ the curves start to superimpose, signalling that equilibrium has been reached. 
At short times, a  similar pattern is displayed by the curves of the subcritical quench (continuous ones) which
stay close to the ones previously discussed. At longer times, for $t>t_{eq}(T_c)$, the correlations move
to the right further than  those of the quench to $T_c$ and there is tendency to accumulate on a curve somewhat
broader than the equilibrium one at $T_c$. This occurs roughly in the range of times $t\in [10^3,10^4]$ which corresponds
to the lifetime of the metastable state. In this time range the correlation is almost time-independent at short $r$, whereas
some evolution can be spotted at large $r$. This signals that the metastable state is not really stable, and
that a prodrome of its failure, which can be interpreted as the build up of critical nuclei, is occurring at large distances.
Finally, roughly for $t>10^4$, a quick growth of correlations is observed, unveiling that metastability is over 
and that the next stage, where ordering takes place, has been entered. In this case, the classification of the 
behaviour of the correlation into three stages (initial evolution, stasis, late evolution) is not sharp because,
as we discuss below, $T$ is relatively far from $T_c$.     

Let us now have a look at the other two panels in Fig.~\ref{fig:corr}. In the central one (b) the quench is closer to $T_c$ and
the accumulation of curves in the metastable state is clearer than in the left panel.
Furthermore, the locus where superposition occurs is also closer to the equilibrium curve at $T_c$
with respect to what is seen in the left panel. 
 In the right panel (c)
the subcritical quench is so close to $T_c$ that the departure from the metastable state cannot even be measured
within the simulated times. Besides that, one sees a pattern similar to the one observed in the other panels,
with the notable difference that the correlations in the metastable state are almost indistinguishable from the
ones of equilibrium at $T_c$. This clarifies our statement that the metastable state is {\it similar} to the equilibrium
one at $T_c$: the closer to $T_c$ the  quench is (from below), the longer the life-time of the metastable state and the
more similar to equilibrium at $T_c$ it is. More precisely, indicating with ${\cal S}_{T}(t)$ a one-time observable
measured in a system quenched to $T\lesssim T_c$, one has $\lim _{T\to T_c^-} {\cal S}_{T}(t)={\cal S}_{eq,T_c}$,
where the latter is the measurement of ${\cal S}$ made in equilibrium at $T_c$. Clearly, this cannot be true 
for any function ${\cal S}$ of the state of the system, particularly if it is weighting hugely the large distance features, 
but the statement is expected to be correct for most quantities of physical interest.

%
%
%
%

The nucleation time $\tau (q,T)$  can be roughly identified as the time when the plateau is over and $R$ 
jumps rather abruptly to a higher value, see Fig.~\ref{fig:e}. This is a violent process corresponding to the 
fast invasion of available space by the super-critical nuclei, as it can be seen from the snapshots 
in Fig.~\ref{fig:q24Snapshot} corresponding to $t=4\cdot 10^4$ and $t=6\cdot 10^4$. It was shown in~\cite{Corberi19}
that $R(t)$ increases exponentially in this time domain. Such speedy relaxation becomes sharper upon increasing $q$.

The duration $\tau (q,T)$ of the metastable state increases as $T\to T_c^-$, as it can be seen in the inset
of Fig.~\ref{fig:e}(a). In the same picture we can also observe that the plateau value $R_m$
is rather $T$-independent while, instead, there is a clear dependence on $q$.
In~\cite{Corberi19} it was found that this value is well approximated by $R_m\propto [\ln (q-4)]^{-1}$
for small $q$ ($q\lesssim 50$) and then, increasing further $q$, it saturates to a value 
of order one. This dependence can be understood recalling that 
\(R_m\) is the inverse energy density excess~(Eq.~[\ref{rvse}]).
The energy density $e_m$ in the metastable state was computed in~\cite{usprevious}
and goes to zero for large \(q\) as $e_m\simeq -a(T) q^{-1/2}$, where 
$a(T)>0$ is a temperature dependent factor.
On the other hand the asymptotic equilibrium quantity 
$e(\infty)<0$ becomes independent on $q$~\cite{Wu97} for large $q$. 
Therefore we conclude that $\Delta e_m=e _m-e (\infty)$ approaches a large-$q$ value from 
below, as observed.
Given the choice of temperatures made in Fig.~\ref{fig:e}(a) (so as to have roughly $\tau (q,T)$ fixed) it is
clear that, working instead at a constant $T$ or even at a finite fraction of $T_c$, i.e. $T=x\cdot T_c$, 
the lifetime of metastability increases with $q$.
Similarly, we have also noticed that the temperature range where metastability occurs widens for larger $q$.
Specifically,  indicating with $T_\ell$ the lower temperature $T$ where metastability is still observed
we find that $(T_c-T_\ell)/T_c$ increases with $q$.

The fast process associated to the steep increase of $R(t)$ ends at a time $t_{coal}$ when 
nuclei of different colour come in contact, a configuration that can be observed in Fig.~\ref{fig:q24Snapshot}
at times $t=8\cdot 10^4$ and/or $t=10^5$ (except for the smallest size $L=80$ where one single colour nucleates,
see the discussion in Sec.~\ref{secmulti}). At this point a coarsening phenomenon sets in 
produced by the competition among domains, as shown in Fig.~\ref{fig:q24Snapshot} and reported
in~\cite{Lifshitz62,Ferrero07,Ibanez07b,Glazier90,Sanders2007,Petri08,Loureiro10,Loureiro12,Olejarz13} and,
for the closely related {\it vector Potts} (or clock) model, in~\cite{Corberi2006}.     
In this kind of evolution one expects dynamical scaling and $R(t)\sim t^{1/2}$, 
which is indeed roughly noted in Fig.~\ref{fig:e} at very long times (for $q=100$ this behavior is likely to be reached
on times longer than the simulated ones). Since this regime is not the focus of this paper we 
content with such a semi-quantitative indication.
 
Before closing this section let us comment on the fact that the data for $R$ 
are free from finite-size effects until domains coarsen up to a length comparable to the system size, which
occurs at a time $t_{fs}(L)$.
This can be seen in Fig.~\ref{fig:e}~(b) where one sees that the curves depart from the
one corresponding to the larger system size ($L=1000$) when $R(t)\simeq 0.1\cdot L$. 
For instance, this occurs at a time $t_{fs}(L=300)$ of order $6\cdot 10^5$, and 
around $t_{fs}(L=160)\simeq 10^5$.
However, the fact that this particular quantity ($R$) does not feel the size
of the system before $t_{fs}(L)$ does not mean that the dynamics is globally free from finite-size effects
in this time domain. Actually, in Fig.~\ref{fig:q24Snapshot} one clearly sees that the configurations 
at a given time look very different in systems of different sizes. The next section is devoted to the discussion 
of this feature.   

\section{Multi-nucleation} 
\label{secmulti}

In the introduction we have briefly discussed  the fact that, at variance with the Ising 
field driven first order transition, in the thermal one the role of the system size is important. The reason is that 
$\tau(q,T)$ and the pseudo-spinodal
temperature depend on system size. In the previous section we have already anticipated that, 
besides this feature, also the nature of the
dynamic escape from the metastable state changes significantly with the system size, as
a visual inspection of Fig.~\ref{fig:q24Snapshot}, where the same quench 
of the $q=24$ model is operated on systems
of different sizes, clearly displays. 
Specifically, 
when a relatively small size is considered 
($L=80$, upper row) 
a single critical droplet forms and grows, invading the whole system (last two snapshots). 
This is what we call mono-nucleation, or 1-nucleation. We have checked that such phenomenon is 
observed rather independently of the thermal realisation of the process.
Notice that the dominating colour (orange) did not seem to be
the favoured one before its outburst (one would have rather predicted the violet 
or yellow colour as better candidates), a fact that shows the rapidity of critical nucleation.   

In the second row the behaviour of a slightly larger system ($L=92$) is shown.
This small size increase is sufficient to modify qualitatively the situation, in that there are now two nucleating colours, the red and the violet ones.
Then, in this case, we have bi-nucleation, or 2-nucleation. Again, this feature is rather
independent of the thermal realisation.
After nucleation, surface tension closes the domain of the minority colour (last snapshot), symmetry is 
definitely broken and equilibrium is attained, but this occurs on much larger times than those of nucleation. 

Increasing further the system size as in the two rows below, the fate of a system changes again
and one observes 3-nucleation and 6-nucleation. 
Notice that, after nucleation, the competition among
domains of different colours leads to a progressive elimination of their number.
Increasing further $L$ (not shown) one can observe $k$-nucleation with $k$ up to $q$ ($q=24$ in this example).

\begin{figure}[h!]
\begin{center}
\hspace{-1.9cm} (a) \hspace{4cm} (b) \hspace{4cm} (c)
\\
\vspace{0.2cm}
\includegraphics[width=1.0\textwidth]{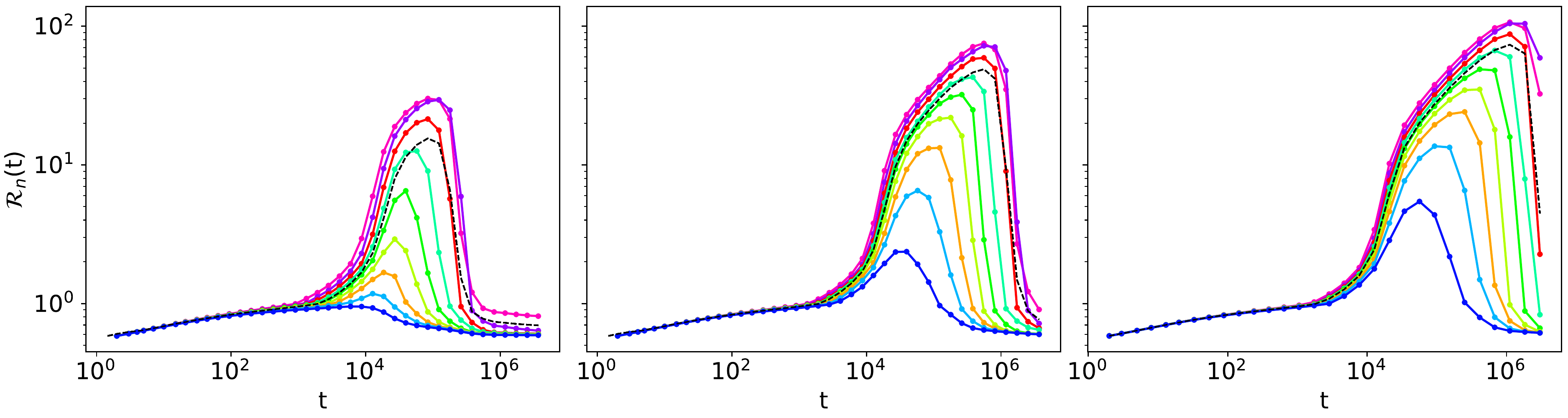}
\end{center}
\vspace{-0.25cm}
    \caption{The typical lengths ${\cal R}_n$ of the various colours are plotted
    against time for the model with $q=9$ quenched to 
    $T=0.9912 \cdot T_c$. The log-log scale is the same for the three panels. 
    The black dashed line is the average ${\cal R}(t)$.
    The system sizes are $L=300$ (a), $L=700$ (b)
	and $L=1000$ (c). Each curve represents an average over $500$ non-equilibrium realisations.
	}
\label{fig_rpart}
\end{figure}

In order to discuss  the multi-nucleation process and its dependence on the system size
at a more quantitative level
we consider the typical lengths of the various colours, ${\cal R}_n(t)$,
which are shown for three different system sizes  in Fig.~\ref{fig_rpart}.
From this figure one clearly sees that ${\cal R}_n$ for the various colours initially
grow, up to a characteristic time $t_n^*$ when they  reach a maximum ${\cal R}^*_n$ and then shrink. 
This means that domains of the corresponding colour grow, reach a maximum size, and then collapse and disappear.
For the prevailing colour (and possibly the next ones in the hierarchy also reaching domain sizes comparable to the system size itself)
denoted by \#1 in Fig.~\ref{fig_rpart}, the interpretation is different, since 
the final decrease of ${\cal R}_1$ must be associated to the invasion of the whole space, according to what
discussed in case ii) at the end of Sec.~\ref{sec:model}. 

\begin{figure}[h!]
\vspace{0.5cm}
\hspace{1.5cm} (a) \hspace{6cm} (b) 
   \begin{center}
    \includegraphics[width=0.45\textwidth]{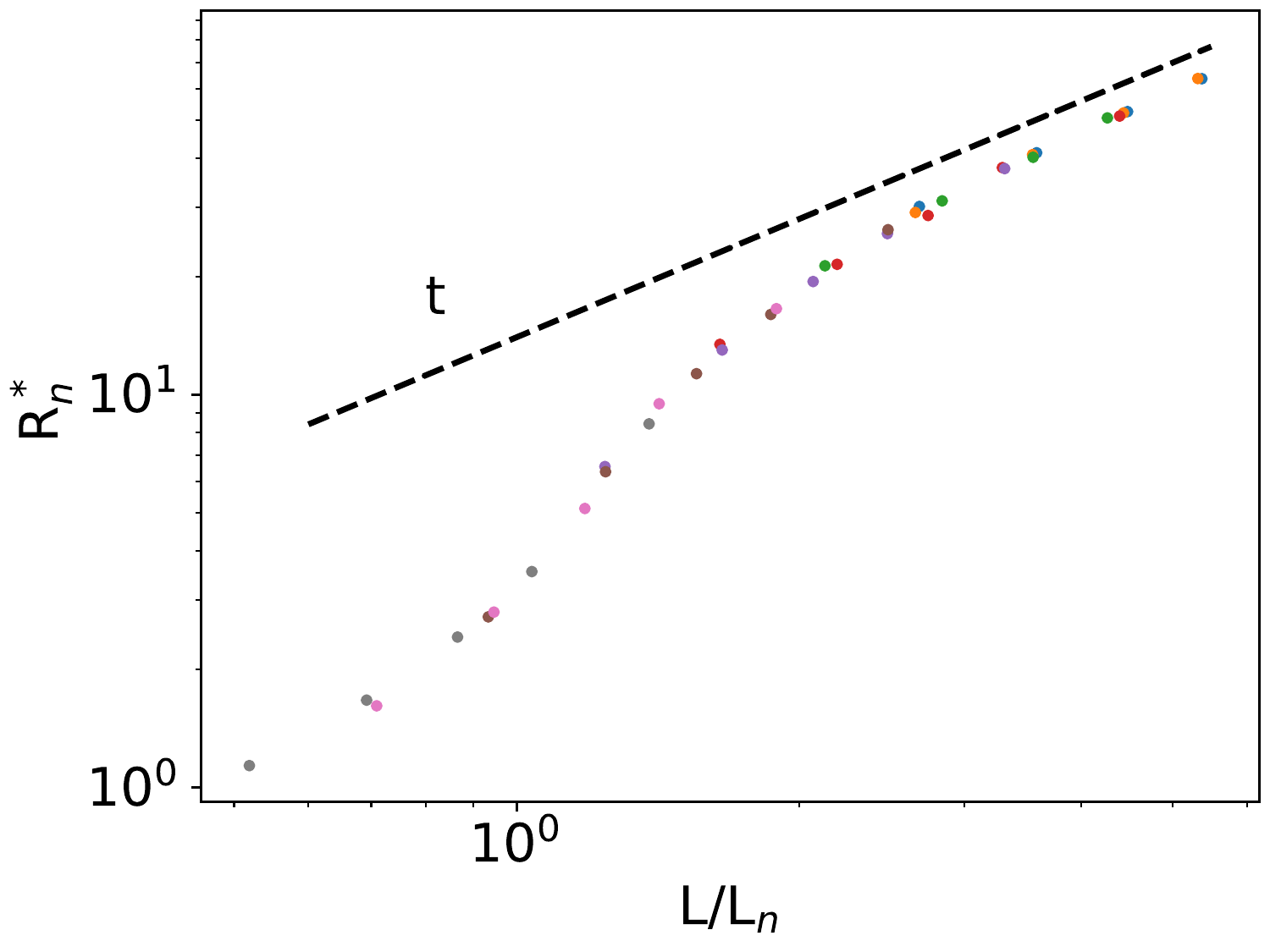}
        \includegraphics[width=0.45\textwidth]{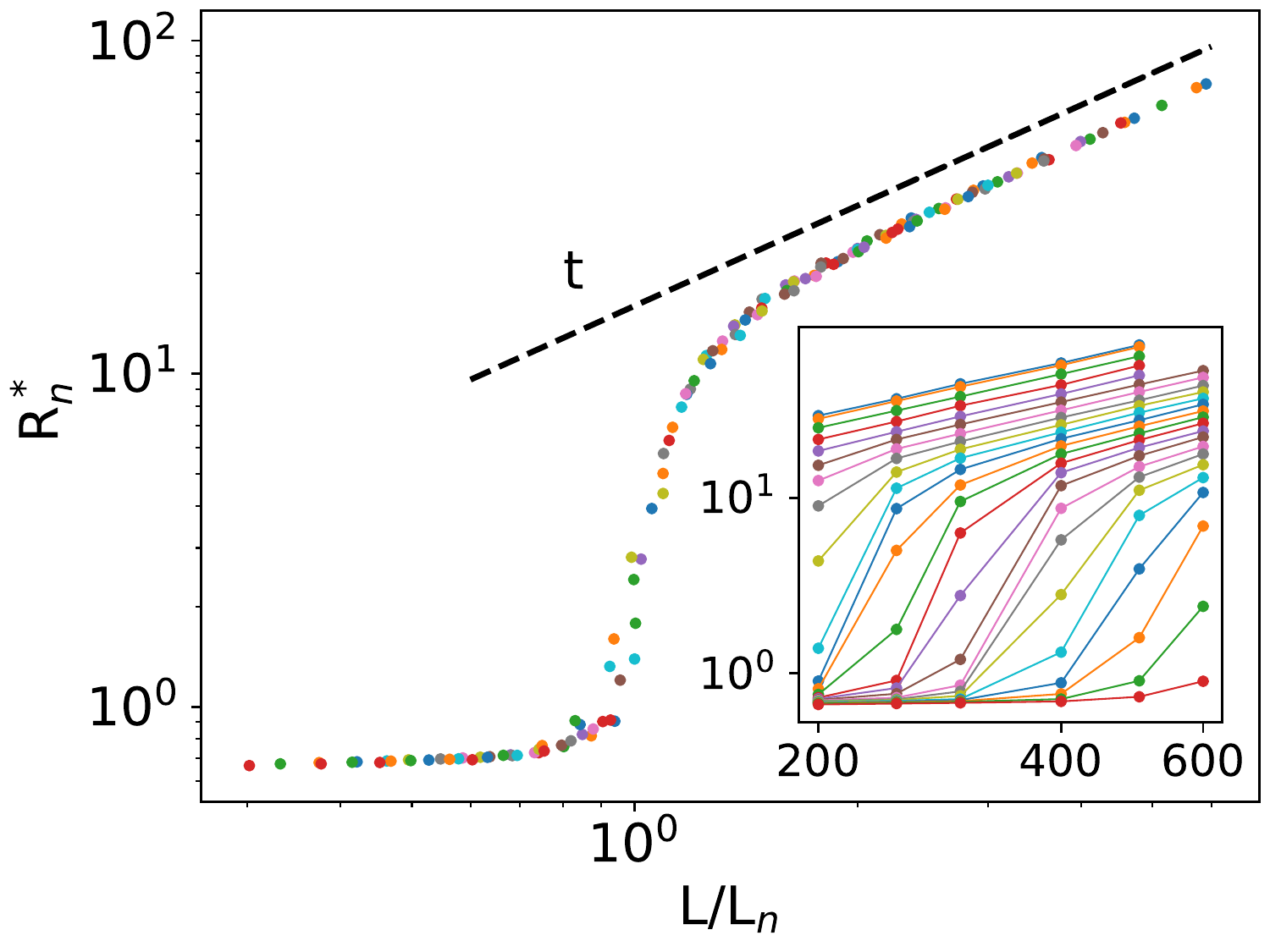}
        \end{center}
\hspace{1.5cm} (c) \hspace{6cm} (d) 
\begin{center}
               \includegraphics[width=0.45\textwidth]{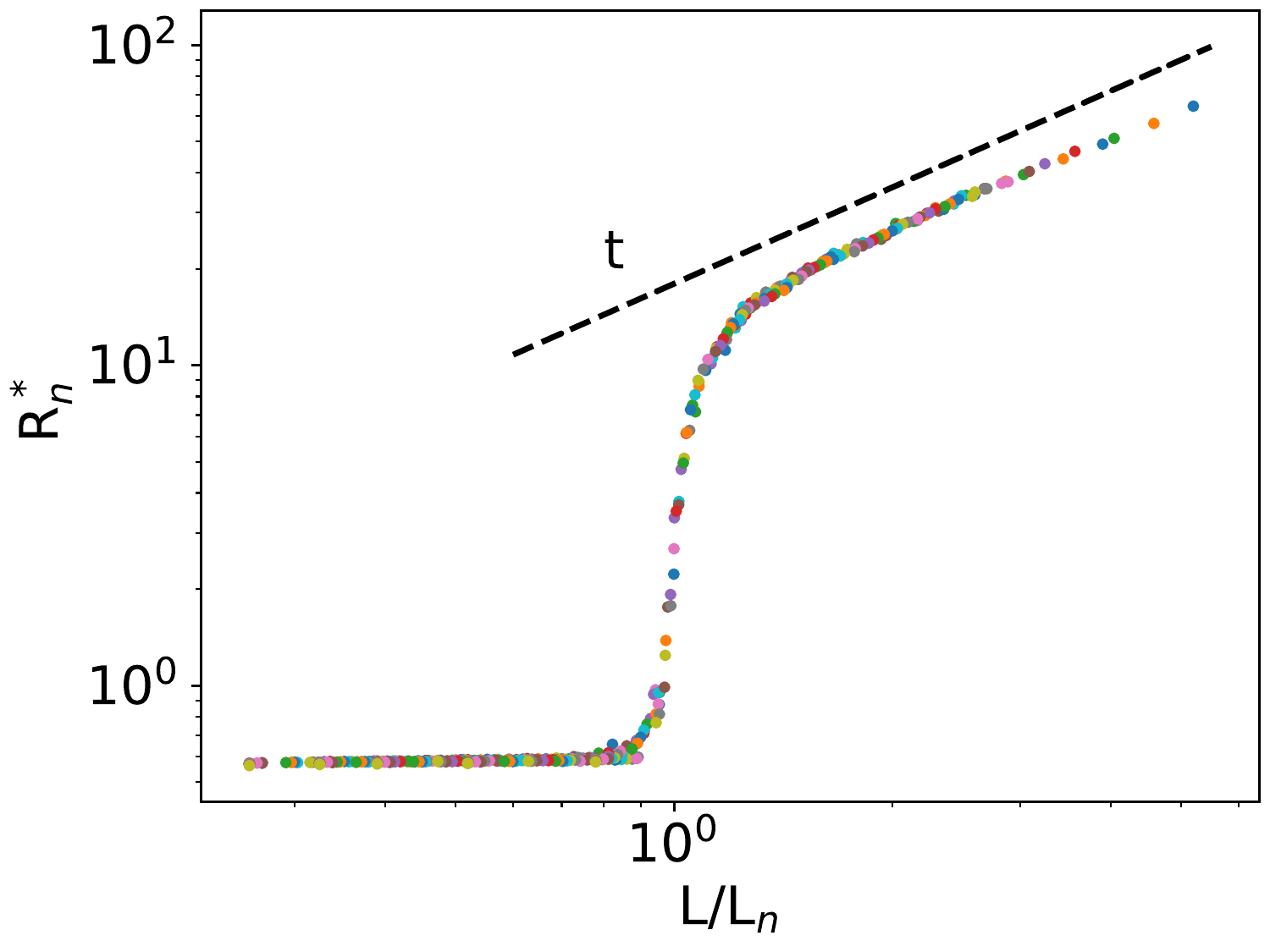}
          \includegraphics[width=0.45\textwidth]{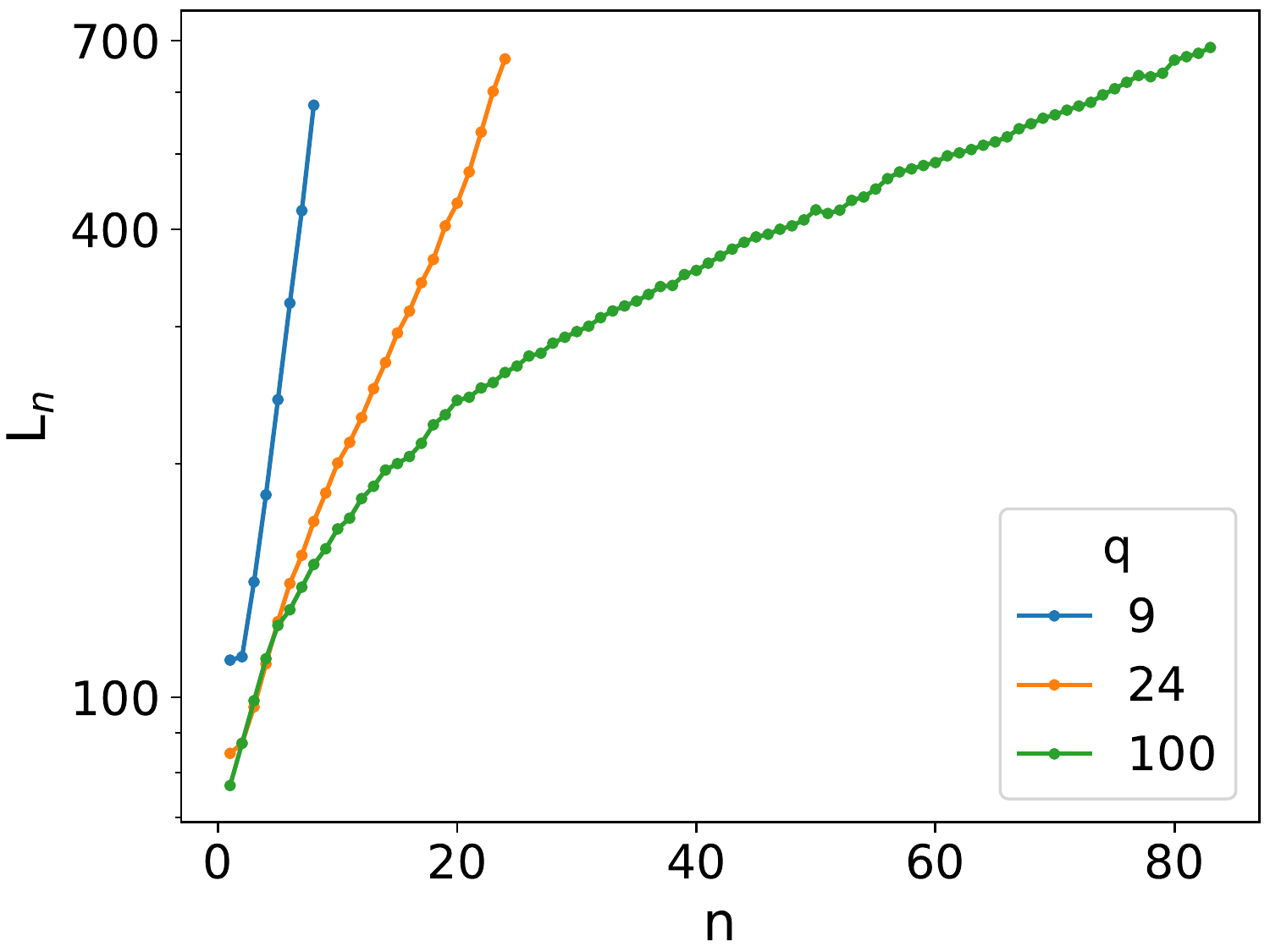}
   \end{center}   
    \caption{The maximum of ${\cal R}_n(q,T,L)$, called 
     ${\cal R}^*_n(q,T,L)$, is plotted in log-log scales against the rescaled size $L/L_n(q,T)$ (see the text) for $q=9$ (a), 
    $q=24$ (b) and $q=100$ (c). The system sizes are  $L=200, \ 250,\ 300,\ 400,\ 500,\ 600, \ 700, \ 800$
    and different colours correspond to different $L$ data.
    The dashed black lines are the linear behavior $y\sim x$. 
    The inset in (b) contains the unscaled
    data (${\cal R}^*_n(q,T,L)$ vs. $L$) for $q=24$ and various values of $n$ (from the most 
    abundant phase on the top, to the less abundant on the bottom). In (d) the dependence of $L_n$ on the colour $n$ is plotted for the
    three values of $q$, in a semilog plot.
    }
\label{fig_peak_vs_L}
\end{figure}

From the comparison between the three different system sizes represented in 
Fig.~\ref{fig_rpart}, one can make some remarks. First,
it is clear that for small sizes some colours do not nucleate, whereas they do in larger systems.
For instance, colour \#9, represented in blue, does not show the exponential increase typical of
nucleation for  $L=300$ and ${\cal R}_9$ does not go beyond $\sim 1$, whereas it does 
grow significantly in the larger systems with $L=700$ and $L=1000$. 
This indicates that ${\cal R}_n$ is a quantity one can use to establish the number of nucleating colours.
Secondly, from Fig.~\ref{fig_rpart} one sees that both the peak time $t_n^*$  and its height ${\cal R}^*_n(q,T,L)$ are increased
in a larger system. Exploiting these two observations we developed a method to determine the number of nucleating
phases as a function of $L$ and $q$, that we now describe. 

First, we measure ${\cal R}_n^*(q,T,L)$ for all $n=1,\dots, q$ and different system sizes. This information is summarised for $q=24$ in 
the inset of  Fig.~\ref{fig_peak_vs_L} (b). 
The first three panels  of this figure, (a)-(c), which refer to different values of $q$, 
show that the data very clearly obey a scaling form
\be
{\cal R}^*_n(q,T,L)=f_q\left [\frac{L}{L_n(q,T)}\right ],
\label{scaling_nucl}
\ee
where $L_n(q,T)$ is a fitting parameter which is plotted 
in Fig.~\ref{fig_peak_vs_L}~(d).
A similar pattern is found for other values of $q$, not reported in the figure.
Notice that imposing collapse of the curves according to Eq.~(\ref{scaling_nucl}) only fixes $L_n(q,T)$ up to an arbitrary
multiplicative constant. We fix it by asking that the steep part of $f_q(x)$ be centred at $x=1$.

\begin{figure}[h]
   \begin{center}
    \includegraphics[width=.8\textwidth]{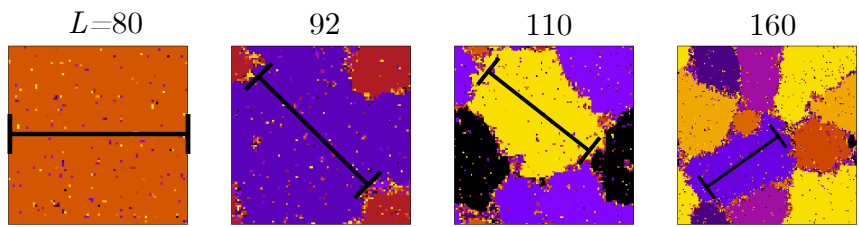}
    \end{center}   
    \caption{Snapshots at \(t=10^5\) of the \(q=24\) Potts model quenched
    to \(T=0.98 \cdot T_c\). This is approximately the time $t_{coal}$ when coarsening, that is competition between touching domains, starts (see Fig.~\ref{fig:q24Snapshot}).
    The black ruler is the length \(L_1 (q,T)\).  One observes $k$-nucleation with $k=1$ for $L=80$, $k=2$ for $L=92$,  
    $k=3$ for $L=110$ and $k=7$ for $L= 160$.}
\label{fig:SnapshotA}
\end{figure}

The scaling function $f_q(x)$ stays small for 
$x\ll 1$, it suddenly increases around $x=1$ (the larger the $q$, the steeper the growth) and it then behaves as 
$f_q(x)\sim x$ for large $x$ (see the dashed lines in Fig.~\ref{fig_peak_vs_L}~(a)-(c)). 
This latter trend indicates that, for each colour such that the representative point lies in this sector, 
the maximum size of the domains is triggered linearly by the system size, as it usually happens for a finite
size effect in a standard (e.g. binary) coarsening system (see also the discussion at the end of this section). Hence, 
we argue that points belonging to this sector of $x$, i.e. $x>1$, correspond to colours that, given the system size $L$, have
nucleated. Analogously, points in the small $x$ sector, i.e. $x<1$, where $f_q(x)$ takes an approximately constant small value,
correspond to colours that did not succeed nucleating. On the basis of this,
we conclude that $x>1$, namely $L>L_n(q,T)$, is the condition
for the $n$-th colour to enter the coarsening stage.
In particular, one can interpret $L_1(q,T)$
as the smallest system size in order to have competition between domains of different colour.
This means that such a size must host the largest domain present at $t_{coal}$ when coarsening starts. Hence we conclude that $L_1 (q,T)$ 
has the physical meaning of the typical size of the largest droplets
in the system at the coalition time $t_{coal}$. This can be appreciated in Fig.~\ref{fig:SnapshotA} where a ruler of length 
$L_1 (q,T)$ is plotted on top of the 
configurations of the model at $t=t_{coal}$.

We also observe that the small-$x$ plateau value of $f_q(x)$ decreases with  increasing $q$. This can be explained
recalling that points laying there are representative of colours that cannot nucleate, and arguing that, as such, their
properties are basically those of the metastable state. The latter, in turn, is akin to the equilibrium state at $T_c$, as discussed,
whose coherence length decreases upon increasing $q$.

\begin{figure}[h!] 
$\;$ \hspace{1.7cm} (a) \hspace{6.3cm} (b)
   \begin{center}
     \includegraphics[width=0.4\textwidth]{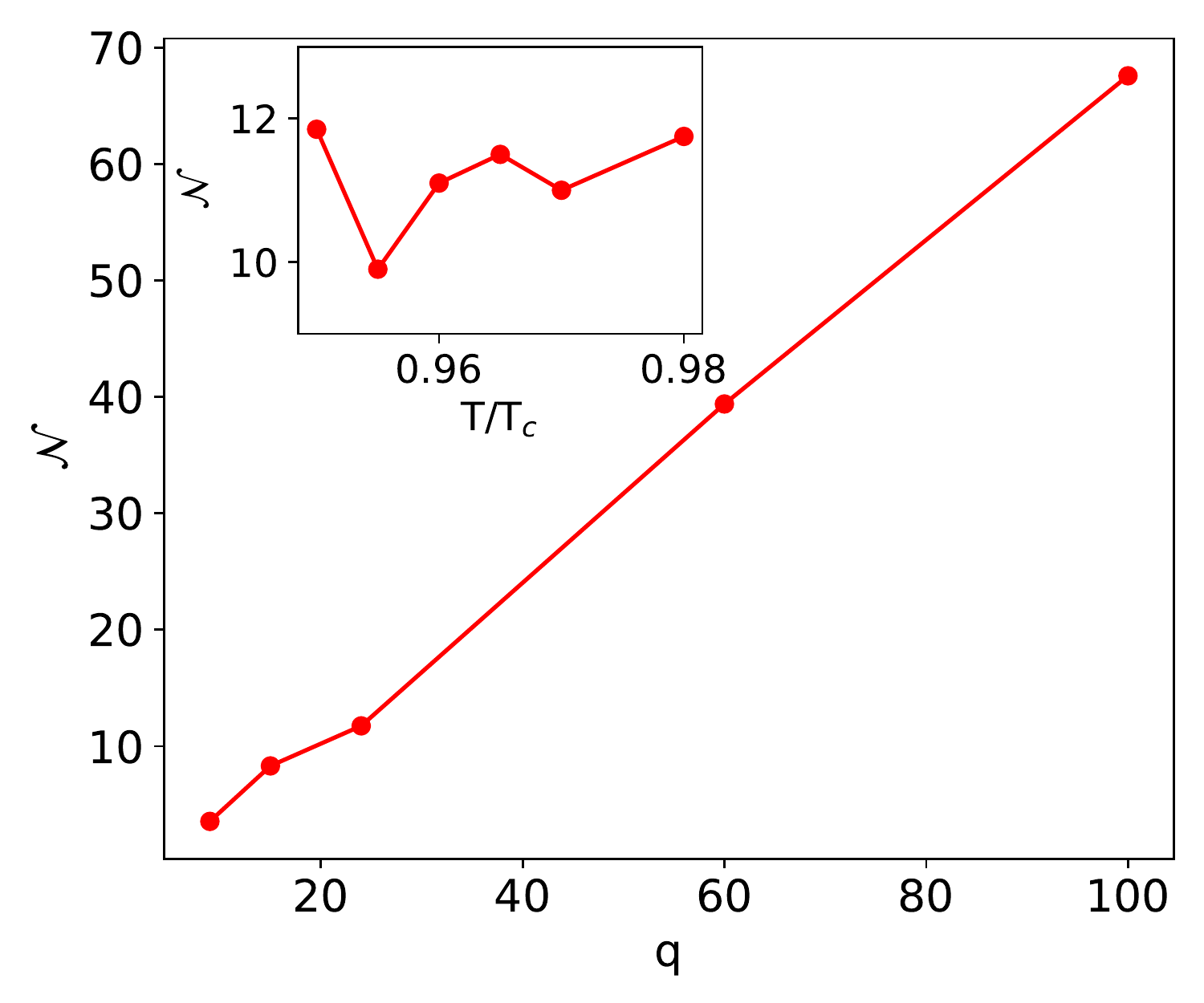}
        \includegraphics[width=0.4\textwidth]{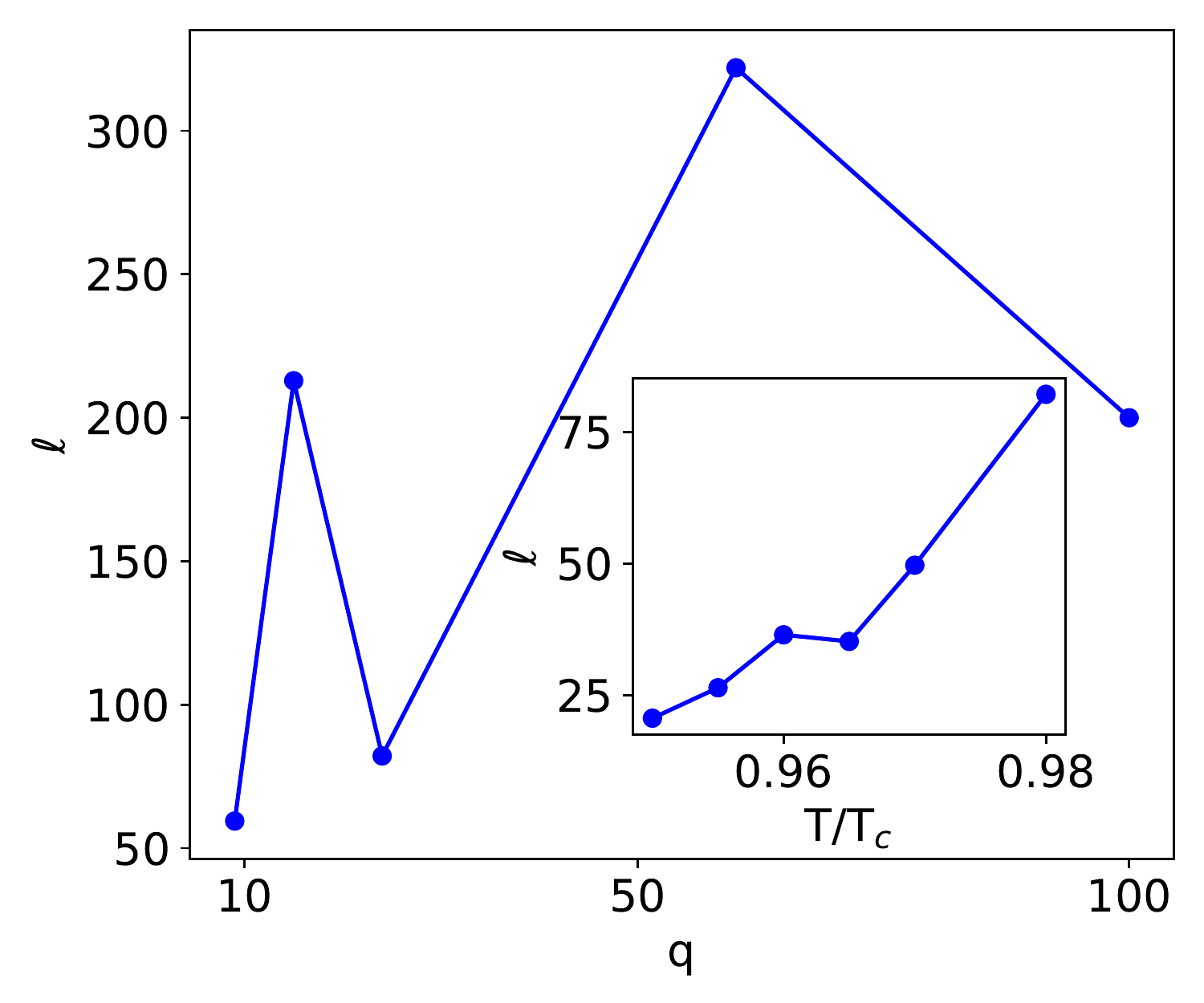}
   \end{center}   
   \vspace{-0.25cm}
    \caption{(a) The dependence of ${\cal N}$ introduced in eq.~(\ref{ellep}), on $q$
     by choosing $T$ at a $q$-dependent value such that
    the lifetime $\tau(q,T)$ of the metastable state is kept constant. In the inset the dependence of ${\cal N}$ on 
    $T/T_c(q)$, for \(q=24\), is displayed, showing that it is not significative (notice the narrow interval on the $y$-axis).
    (b) The dependence of $\ell$, also defined in eq.~(\ref{ellep}), on $q$ with the same choice of a $q$-dependent $T$  as in (a). 
    In the inset, the variation of $\ell$ as a function of $T/T_c(q)$ for \(q=24\).}
\label{fig:APdependences}
\end{figure}

Looking at the plot of $L_n(q,T)$ in  Fig.~\ref{fig_peak_vs_L}~(d), one finds an exponential dependence
\be
L_n(q,T)=\ell (q,T) \,e^{n/{\cal N}(q)}.
\label{ellep}
\ee
This law describes well all the data for $q=9,24$, and the large $n$ region for $q=100$
(with fitting parameters $\ell (q,T)=59.43$, 82.97, 200.00 and ${\cal N}(q)=3.546$, 11.848, 67.568  for $q=9,24,100$, respectively). 
Furthermore, we found an analogous behaviour for other values of $q$ not portrayed in Fig.~\ref{fig_peak_vs_L}. 
Repeating the procedure above for different temperatures we find that ${\cal N}$ is, within errors, independent of $T$.
This is shown in the inset of Fig.~\ref{fig:APdependences}~(a) for a fixed value of $q$, $q=24$ in this case. 
It seems that the behaviour reported can be ascribed more to some random effect, or errors, rather than to a genuine
dependence (notice also that the fluctuations of ${\cal N}$ are only around 10\% of ${\cal N}$ itself). 
Similar results are found for 
other values of $q$. This explains the parameter dependencies written in Eq.~(\ref{ellep}).
 
As shown in Fig.~\ref{fig:APdependences}~(a)  ${\cal N}(q)$ turns out  to increase in a roughly 
linear way with $q$. In this case, at variance with the $T$ dependence discussed before, the variation of ${\cal N}(q)$
is monotonic and important, as compared to ${\cal N}(q)$ itself. 
Moving to $\ell (q,T)$, its dependence on $T/T_c(q)$ (at fixed $q=24$) can be appreciated in 
inset of the right panel of Fig.~\ref{fig:APdependences}, where one finds that $\ell (q,T)$ increases in a roughly linear way with $T$. 
On the other hand, with the choice of $T$ made in our simulations, which as discussed earlier
correspond to fix the lifetime $\tau (q,T)$ of the metastable state, the dependence of $\ell (q,T)$ on $q$, 
shown in the right panel of Fig.~\ref{fig:APdependences}, cannot be straightforwardly interpreted.

We now discuss the meaning of the length $\ell (q,T)$ appearing in Eq.~(\ref{ellep}).
We consider an intermediate range of values of $q$ such that the following two conditions are obeyed: 
a) $q$ is sufficiently small as to have the exponential form~(\ref{ellep}) 
valid down to $n=1$ (as already commented, this is true for $q=9,24$ but not for $q=100$) and b)
${\cal N}(q)\gg 1$. For example, the two conditions are reasonably obeyed for $q=24$, since ${\cal N}(q=24)=11.848 \gg 1$.
In such an intermediate range of $q$, after Eq.~(\ref{ellep}) one has 
$\ell (q,T)\simeq L_1(q,T)$, the typical size associated to the prevailing colour whose interpretation 
was discussed above. Hence $\ell (q,T)$ amounts to the typical size of the largest droplets
in the system at the coalition time $t_{coal}$. For $q$ outside such intermediate range, e.g. for large $q$, we expect
$\ell (q,T)$ to have a related interpretation, even though we cannot derive it directly from Eq.~(\ref{ellep}). 
Let us also observe that from Fig.~\ref{fig:APdependences} one extrapolates
$\ell (q,T)\to \infty$ for $T\to T_c$, implying that the metastable state becomes the stable equilibrium one, as expected. 

Coming to the issue of multi-nucleation, applying the argument above to $L_n(q,T)$, we can say that the $n$-th colour has nucleated if $L>L_n$. 
Using Eq.~(\ref{ellep}) we arrive at the conclusion that, given a system size $L$, one has $k$-nucleation 
with 
\be
k(q,T,L)=\left \{{\cal N}(q)\ln \left [\frac {L}{\ell (q,T)}\right ]\right \}_{\in [1,q]},
\label{num_phases}
\ee
where the notation $\{Z\}_{\in [1,q]}=\max \{1,\min\{q,Z\}\}$ simply accounts for the constraint $k\in [1,q]$.
One can easily check that, using the values of ${\cal N}(q)$ and $\ell (q,T)$ in this equation, one can correctly predict
the number of nucleating colours observed in the various cases of Fig.~\ref{fig:q24Snapshot}.
Notice however that the domain of validity of this equation is inherited by that of Eq.~(\ref{ellep}), hence if $q$ is large
it holds true only for sufficiently large values of $k$. 

This result shows that, for a given $q$, the number of nucleating phases grows only logarithmically with the system size.
When the thermodynamic limit is taken from the onset, all phases nucleate.
Also, if the size of the system is large enough, i.e. $L\gg \ell (q,T)$, 
increasing $q$ reduces the number of nucleating phases. 
Let us also notice that Eq.~(\ref{num_phases}) informs us on the number of nucleating phases, but does not predict
the time needed to nucleate. Clearly, if this time grew towards infinity, no nucleation would be found. 
In the large $q$-limit, it was also observed that a  different mechanism  takes over, see for instance Fig. 6(c)(d) in~\cite{ChCuPi21}, with, possibly, 
a different scaling of $k$ with $L$ that still needs to be investigated in depth.

The discussion above shows that the multi-nucleation dynamics of the two dimensional Potts model  
is interested by peculiar finite size effects, rather different from the usual ones arising in the phase-ordering of 
binary systems. 
In the latter,  there are only two competing colours and the growth of the domains does not feel the 
finiteness of the system until some time $t_{end}(L)$, when the domains' size becomes comparable to 
$L$. At this point the 
coarsening process is interrupted and the system enters a final stage whereby equilibrium is approached. 
For times $t\ll t_{end}$ any (intensive) measurement does not depend on the value of $L$. 
In the Potts model, one observes an analogous behaviour when observing a quantity like $R$, as we discussed at the end
of Sec.~\ref{theprocess}. This is also true also for the sizes ${\cal R}_1$ and ${\cal R}_2$ of the two prevailing colours.
However, the number $k$ of coarsening phases depends on $L$ 
up to a value of $L$ as large as 
$L_q$, given in Eq.~(\ref{ellep}). Notice that this characteristic size diverges both as $T\to T_c$ for 
a given $q$ and as $q\to \infty$ for a given $T$, as discussed above. 
Worst,  the sizes ${\cal R}_n$ ($n\in [3,q]$)  of 
the domains of the minority colours, do depend on $L$ even if measured at times $t\ll t_{end}$
(see Fig.~\ref{fig_rpart}).
This strange behaviour is perhaps at the origin of many controversies on the size dependence of the metastable dynamics.

\begin{figure}[h!]
\begin{center}
\includegraphics[width=0.45\textwidth]{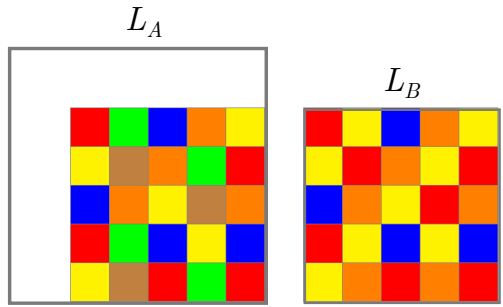}
\end{center}
\caption{Two pictorial configurations of the bidimensional Potts model on the square lattice. The system on the right has a size $L_B$ and 
the picture represents the whole lattice. The one on the
left has a larger size $L_A>L_B$ but in the picture only a part of it, of size $L_B$, is shown.
The two systems are characterised by the same size of the domains,
but in the right one the green and brown phases have disappeared.}
\label{fig_schematic}
\end{figure}

One might wonder how it may be possible that the unrestricted quantity $R$ be size-independent 
whereas the restricted ones ${\cal R}_n$ do depend on $L$, given that $R$ is morally the average of the ${\cal R}_n$s.
This occurs because $R$ only fixes the average size of domains but not their colour, as it is
schematically illustrated in Fig.~\ref{fig_schematic}.
The configuration on the left shows a portion of a system of size $L_A$ at a time $t$ 
where six phases are present. On the right, instead, an analogous system with a different size 
$L_B< L_A$ is represented at the same time $t$. The represented portion of the first
system equals the total size of the second. It can be seen that, although two phases 
(the green and the brown one) are absent in the smaller lattice (hence the number of colours is $L$ dependent), the size of the domains
does not change (hence is $L$ independent).  

\vspace{0.5cm}

\section{Discussion}
\label{sec:discussion}

In this paper we studied the multi-nucleation phenomenon after a shallow quench
of the Potts model with $4<q\leq 100$  to $T\lesssim T_c(q)$. A related study
of the model with $q\gg 100$ addressing deep quenches to $T\ll T_c(q)$ is presented
in a companion article by some of us~\cite{ChCuPi21}.  

The main result of this paper, which is contained in Eq.~(\ref{num_phases}), is that the number of 
nucleating phases of the two-dimensional Potts model with not too large $q$ 
increases logarithmically with the system size.
The logarithmic behaviour informs us that this feature cannot be related to the trivial geometrical fact
that a larger system can {\it accommodate} a larger number of developing colours. If this were the origin, 
the snapshots in Fig.~\ref{fig:q24Snapshot} should display an equal number of nucleating phases in 
portions with equal area of systems with different sizes.
This, however, is not observed. For instance, 
comparing the systems of relative double size $L=80$ (first line) and $L=160$ (last row), at time
$t=80000$, one sees that there is only one nucleating colour (orange, besides a remnant of the disordered
phase) in the smallest system,  whereas there are typically more than one colour in a portion of area $80\times 80$
of the system with $L=160$. This raises an important 
question which, by the way, concerns also the size dependence
of the metastable state recalled in the introduction: how can a system with a very short coherence length 
(of the order of the nuclei size, as it can also be appreciated from the correlation function of Fig.~\ref{fig:corr}) feel the boundary effects?  
Rephrased differently, if nucleation is a local process, since correlations are short, how could it be influenced
by a global property such as the total size? 

Here we provide a possible, although at the moment completely speculative, answer. In a series of 
papers~\cite{Blanchard14,Disordered,Diluted} some of us showed that in a rather broad class of $2d$ binary models
the fate of the system, namely which of the two colours will eventually invade the sample,
is decided at a very early stage when a percolation cluster grows as large as the whole sample. 
Since percolation is an uncorrelated phenomenon, this occurs when spatial correlations are still 
minimal and the ferromagnetic domains microscopic. 
However, such tenuous invisible structure determines the phase that will eventually
(at much longer times) win the competition simply by touching the boundaries. 
Therefore, it may be conjectured that something similar -- in a way to be better investigated --
also happens in the present case with $q$ colours. Namely, a large but slender structure (not necessarily
with a critical percolation topology as in the case with $q=2$), which cannot be easily associated to
the pattern of the nucleating domains, could invade the sample till its boundaries, and determine the  
fate of the colours. This seems to agree with the observation, made regarding Fig.~\ref{fig:corr}, 
that within the metastable state $C(r,t)$ does not vary in time at small $r$, whereas it does so at large distances.
Which are the geometrical features to look for to identify such large scale structure, if any, 
is still an open  question. 

The mechanism for ordering in systems with $q\gg 100$ is rather different from the one discussed here. In
these systems after an extremely long waiting period (the longer the closer to $T_c(q)$ at fixed $q$, 
or the larger $q$ at $T/T_c(q)$ fixed) the systems suddenly jump to
a partially ordered state with large and blocked domains. No ``sand'' 
(disordered regions between clusters that can be seen in Fig.~\ref{fig:q24Snapshot})
lubricates the motion of the domain walls in this case and the 
interfaces are mostly straight. Although the quench is to high $T$, even close to 
$T_c(q)$, these blocked states resemble the ones at zero temperature. We postpone the analysis of the ordering process
under these ($q, T/T_c(q))$ conditions to a future publication. Still, an idea of the way in 
which the typical length scale $R$ increases in time in this regime of parameters can be reached by looking at Fig. 6~(c)-(d) 
in~\cite{ChCuPi21}.

\vspace{0.25cm}

\noindent
{\bf Acknowledgements}
We thank A. Vezzani for very useful comments, and F. Chippari with whom two of us performed a parallel study of quenches 
towards parameters farther away from $T_c(q)$~\cite{ChCuPi21}.

\end{document}